\documentclass[]{emulateapj}
\usepackage{graphicx}
\usepackage{graphics}
\usepackage{float}

\usepackage{natbib}
\usepackage{txfonts}
\bibpunct{(}{)}{;}{a}{,}{,~}

\shorttitle{The effect of dust cooling in the fragmentation of star-forming clouds}
\shortauthors{Dopcke et al.}

\begin{document}

\title{The effect of dust cooling on low-metallicity star-forming clouds}

\author{Gustavo Dopcke, Simon C. O. Glover, Paul C. Clark and Ralf S. Klessen}
\affil{Zentrum f\"{u}r Astronomie der Universit\"{a}t Heidelberg, Institut f\"{u}r Theoretische Astrophysik, Albert-Ueberle-Str. 2, 69120 Heidelberg, Germany}

\begin{abstract}
The theory for the formation of the first population of stars (Pop III) predicts a IMF composed predominantly of high-mass stars, in contrast to the present-day IMF, which tends to yield stars with masses less than 1 M$_{\odot}$. The leading theory for the transition in the characteristic stellar mass predicts that the cause is the extra cooling provided by increasing metallicity and in particular the cooling provided at high densities by dust. The aim of this work is to test whether dust cooling can lead to fragmentation and be responsible for this transition. To investigate this, we make use of high-resolution hydrodynamic simulations.
We follow the thermodynamic evolution of the gas by solving the full thermal energy equation, and also track the evolution of the dust temperature and the chemical evolution of the gas. We model clouds with different metallicities, and determine the properties of the cloud at the point at which it undergoes gravitational fragmentation. We follow the further collapse to scales of an AU when we replace very dense, gravitationally bound, and collapsing regions by a simple and nongaseous object, a sink particle.
Our results suggest that for metallicities as small as 10$^{-5} \rm Z_{\odot}$, dust cooling produces low-mass fragments and hence can potentially enable the formation of low mass stars.
We conclude that dust cooling affects the fragmentation of low-metallicity gas clouds and plays an important role in shaping the stellar IMF even at these very low metallicities.
\end{abstract}


\section{Introduction}\label{int}
The first burst of star formation in the Universe is thought to give rise to massive stars, the so-called Population III, with current theory predicting masses in the range 20-150 M$_{\odot}$ \citep{2002Sci...295...93A, 2002ApJ...564...23B, 2007ApJ...654...66O, 2008Sci...321..669Y}.
This contrasts with present-day star formation, which tends to yield stars with masses less than 1 M$_{\odot}$ \citep{2002Sci...295...82K, 2003PASP..115..763C}, and so at some point in the evolution of the Universe there must have been a transition from primordial (Pop. III) star formation to the mode of star formation we see today (Pop. II/I).\\
When gas collapses to form stars, gravitational energy is transformed to thermal energy and unless this can be dissipated in some fashion, it will eventually halt the collapse. Thermal energy can be dissipated by processes such as atomic fine structure line emission, molecular rotational or vibrational line emission, or the heating of dust grains. In some cases, these processes are able to cool the gas significantly during the collapse. This temperature drop can promote gravitational fragmentation \citep{2004RvMP...76..125M, 2007prpl.conf..149B} by diminishing the Jeans mass, which means that instead of forming very massive clumps, with fragment masses corresponding to the initial Jeans mass in the cloud, it can instead form even more fragments with lower masses.\\  
If the gas is cooled only by molecular hydrogen emission, numerical simulations show that the stars should be very massive \citep{2002Sci...295...93A, 2002ApJ...564...23B, 2007ApJ...654...66O, 2008Sci...321..669Y}.
This happens because the H$_2$ cooling becomes inefficient for temperatures bellow 200K and densities above $10^4 \rm cm^{-3}$. At this temperature and density, the mean Jeans mass at cloud fragmentation is 1,000 times larger than in present-day molecular clouds,
\begin{equation}\label{BEm}
M_{\rm frag} \approx 700M_{\odot} \left(\frac{T_{\rm frag}}{200 \rm K}\right)^{3/2} \left(\frac{n_{\rm frag}}{10^4 \rm cm^{-3}}\right)^{-1/2},
\end{equation}
for an atomic gas with temperature $T_{\rm frag}$ and number density $n_{\rm frag}$.\\
Metal line cooling and dust cooling are effective at lower temperatures and larger densities, and so the most widely accepted cause for the transition from Pop. III to Pop. II is metal enrichment of the interstellar medium by the previous generations of stars. This suggests that there might be a critical metallicity Z$_{\rm crit}$ at which the mode of star formation changes.\\ 
The main coolants that have been studied in the literature are CII and OI fine structure emission \citep{2001MNRAS.328..969B, 2003Natur.425..812B,  2006ApJ...643...26S, 2007MNRAS.380L..40F}, and dust emission. C and O are identified as the key species because in the temperature and density conditions that characterise the early phases of Pop. III star formation, the OI and CII fine-structure lines dominate over all other metal transitions \citep{1989ApJ...342..306H}. By equating the CII or OI fine structure cooling rate to the compressional heating rate due to free-fall collapse, one can define critical abundances $[\rm C/H] = -3.5$ and $[ \rm O/H] = -3.0$\footnote{$[\rm X/ \rm Y] = log_{10}(N_{\rm X}/N_{\rm Y})_{\star} - log_{10}(N_{\rm X}/N_{\rm Y})_{\odot}$, for elements X and Y, where $\star$ denotes the gas in question, and where $\rm N_X$ and $\rm N_Y$ are the mass fractions of the elements X and Y.} for efficient metal line cooling \citep{2003Natur.425..812B}.
However, previous works \citep{2009ApJ...696.1065J, 2009ApJ...694.1161J} show that this metallicity threshold does not represent a critical metallicity: the fact that metal-line cooling has a larger value than the compressional heating does not necessarily lead to fragmentation.\\
Dust-cooling models predict a much lower critical metallicity $(\rm Z_{\rm crit} \approx 10^{-5} \rm Z_{\odot})$.
The conditions for fragmentation in the low-metallicity dust cooling model are predicted to occur in high density gas, where the distances between the fragments can be very small \citep{2000ApJ...534..809O, 2005ApJ...626..627O, 2002ApJ...571...30S, 2006MNRAS.369.1437S, 2010MNRAS.402..429S}. In this regime, interactions between fragments will be common, and analytic models of fragmentation are unable to predict the mass distribution of the fragments. A full 3D treatment, following the fragments, is needed.\\
Initial attempts were made by \cite{2006ApJ...642L..61T,2008ApJ...676L..45T} and \cite{2008ApJ...672..757C}. However, these treatments used a tabulated equation of state, based on results from previous one-zone chemical models \citep{2005ApJ...626..627O}, to determine the thermal energy. This approximation assumes that the gas temperature adjusts instantaneously to a new equilibrium temperature whenever the density changes and hence ignores thermal inertia effects. This may yield too much fragmentation.\\
In this work, we improve upon these previous treatments by solving the full thermal energy equation, and calculating the dust temperature through the energy equilibrium equation. We assume currently that the only significant external heat source is the CMB, and include its effects in the calculation of the dust temperature.\\

\section{Simulations}

\subsection {Numerical method}\label{numap}
We model the collapse of a low-metallicity gas cloud using a modified version of the Gadget 2 \citep{2005MNRAS.364.1105S} smoothed particle hydrodynamics (SPH) code. To enable us to continue our simulation beyond the formation of the first very high density protostellar core, we use a sink particle approach \citep{1995MNRAS.277..362B}, based on the implementation of \cite{2005A&A...435..611J}. Sink particles are created once the SPH particles are bound, collapsing, and within an accretion radius, $h_{acc}$, which is taken to be 1.0 AU. The threshold number density for sink particle creation is $5.0 \times 10^{13} \rm cm^{-3}$. At the threshold density, the Jeans length at the minimum temperature reached by the gas is approximately one AU, while at higher densities the gas becomes optically thick and begins to heat up. Further fragmentation on scales smaller than the sink particle scale is therefore unlikely to occur. For further discussion see \cite{2011ApJ...727..110C}.\\
To treat the chemistry and thermal balance of the gas, we use the same approach as in \citet{2011ApJ...727..110C}, with two additions: the inclusion of the effects of dust cooling, as described below, and formation of H2 on the surface of dust grains \cite[see][]{1979ApJS...41..555H}. The \citet{2011ApJ...727..110C} chemical network and cooling function were designed for treating primordial gas and do not include the chemistry of metals such as carbon or oxygen, or the effects of cooling from these atoms, or molecules containing them such as CO or H$_{2}$O. We justify this approximation by noting that previous studies of very low-metallicity gas \citep[e.g.][]{2005ApJ...626..627O, 2010ApJ...722.1793O} find that gas-phase metals have little influence on the thermal state of the gas. \cite{2010ApJ...722.1793O} showed that H2O and OH are efficient coolants at $10^8 < n < 10^{10} \rm cm^{-3}$ for their one-zone model. In their hydrodynamical calculations, however, the collapse is faster, and the effect of H2O and OH is not perceptible. Therefore we do not expect oxygen-bearing molecules to have a big effect on the thermal evolution of the gas. For the metallicities and dust-to-gas ratios considered in this study, the dominant sources of cooling are the standard primordial coolants (H$_{2}$ bound-bound emission and collision-induced emission) and energy transfer from the gas to the dust.
\begin{figure}[H]
  \centering
    \includegraphics[width=0.5\textwidth]{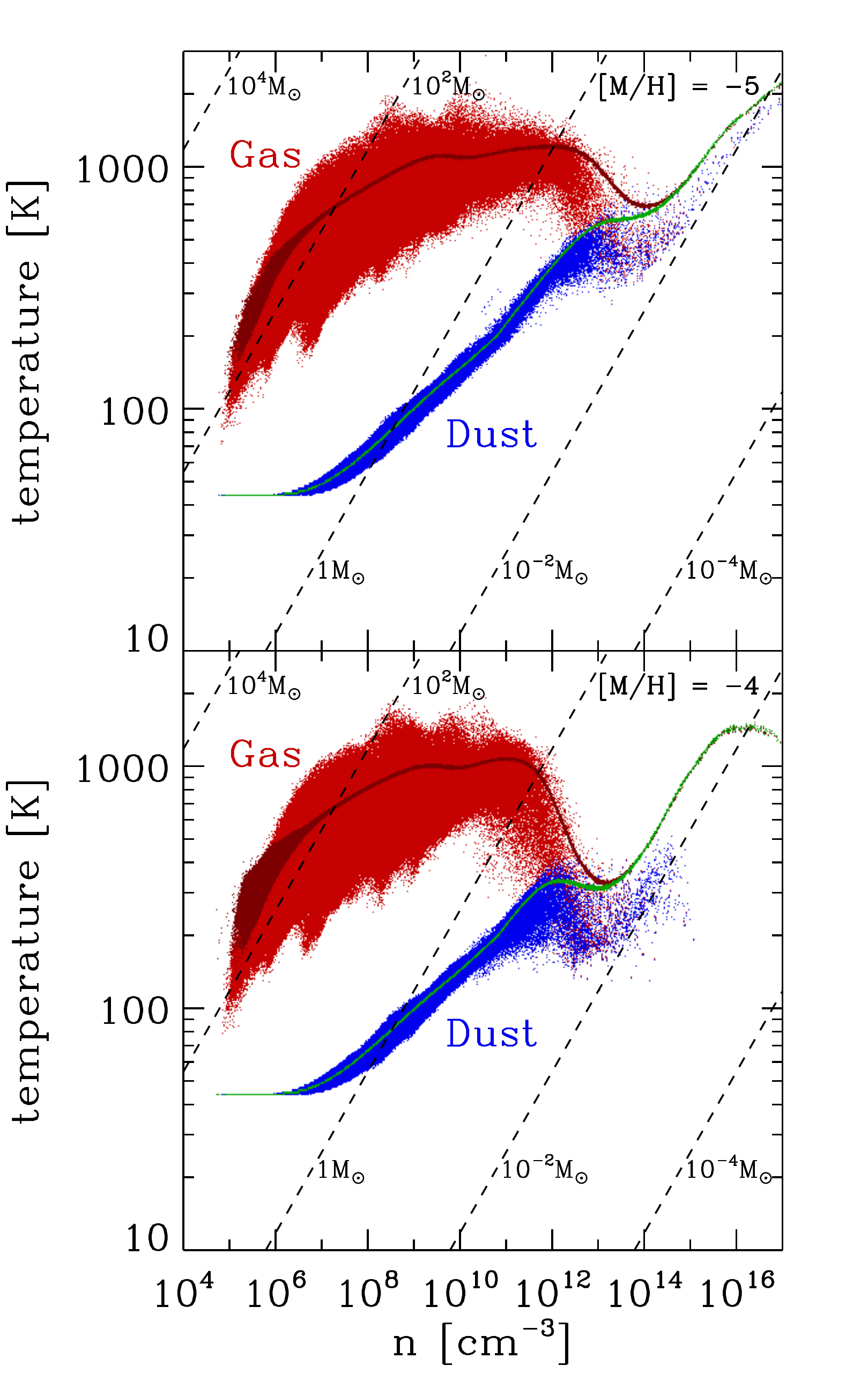}
  \caption{Results of our low-resolution simulations, showing the dependence of gas and dust temperatures on gas density for metallicities $10^{-4}$ and $10^{-5}$ times the solar value. In red, we show the gas temperature, and in blue the dust temperature for the turbulent and rotating cloud. The simple core collapse is overploted in dark red and green. The points with thinner features are from the simulations without rotation or turbulence, while those showing more scatter come from the simulations with rotation and turbulence. The dashed lines show constant Jeans mass values.  
 \label{nt}}
\end{figure}

\subsubsection{Dust cooling}
Collisions between gas particles and dust grains can transfer energy from the gas to the dust (if the gas temperature $T$ is
greater than the dust temperature $T_{\rm gr}$), or from the dust to the gas (if $T_{\rm gr} > T$). The rate at which energy is
transferred from gas to dust is given by \citep{1979ApJS...41..555H}
\begin{equation}\label{gascool}
\Lambda_{\rm gr} = n_{\rm gr} n \bar{\sigma}_{\rm gr} v_{\rm p}f(2kT - 2kT_{\rm gr}) \: \: {\rm erg} \: {\rm s^{-1}} \:
{\rm cm^{-3}},
\end{equation}
where $n_{\rm gr}$ is the number density of dust grains, $n$ is the number density of hydrogen nuclei, $\bar{\sigma}_{\rm gr}$ is the mean dust grain cross-section, $v_{\rm p}$ is the thermal speed of the proton, and $f$ is a factor accounting for the ontribution of species other than protons, as well as for charge and accommodation effects. We assume that $\bar{\sigma}_{\rm gr}$ is the same as for Milky Way dust, and that the number density of dust grains is a factor ${\rm Z}/{\rm Z_{\odot}}$ smaller than the Milky Way value. 
To compute the rate at which the dust grains radiate away energy, we use the approximation
\citep{2007A&A...475...37S}
\begin{equation}\label{radheat}
\Lambda_{\rm rad} = 4\sigma_{\rm sb} n_{\rm gr} \frac{(T_{\rm gr}^4 - T_{\rm cmb}^4)} {\Sigma^2 \kappa_R + \kappa_{P}^{-1}},
\end{equation}
where $T_{\rm cmb}$ is the CMB temperature, $\sigma_{\rm sb}$ is the Stefan-Boltzmann constant, $\kappa_{P}$ and $\kappa_{R}$ are the Planck and Rosseland mean opacities and $\Sigma$ is the column density of gas measured along a radial ray from the particle to the edge of the cloud.\\
As explained by \cite{2007A&A...475...37S}, this expression has the correct behaviour in the optically thin and optically thick limits, and interpolates between these two limits in a smooth fashion. In practice, 
we approximate further by assuming that the Planck and Rosseland mean opacities are equal
and by using the fact that $\Sigma \sim \rho L_{\rm J}$ for a gravitationally collapsing gas, where $\rho$ is the mass density of the gas, and $L_{\rm J}$ is the Jeans length, given by $L_{\rm J} = (\pi c_{s}^{2} / G \rho)^{1/2}$, where $c_{s}$ is the speed of sound in the gas. By approximating $\Sigma$ in this fashion, we avoid the computational difficulties involved with measuring column densities directly in the simulation, while still following the behaviour of the gas reasonably accurately in the optically thick regime. In any case, most of the interesting behaviour that we find in our simulations occurs while dust cooling remains in the optically thin regime. To compute the temperature of the dust grains, we assume that the dust is in thermal equilibrium, and hence solve the equilibrium equation
\begin{equation}\label{teq}
\Lambda_{\rm gr} - \Lambda_{\rm rad} = 0.
\end{equation}
This equation is transcendental, so we solve it numerically.

\subsubsection{Dust opacity}\label{opac}
We follow the dust opacity model of  \cite{2001ApJ...557..736G}, and we calculate the opacity as a function of the dust temperature in the same fashion as in \citet{2006MNRAS.373.1091B}. To convert from the frequency-dependent opacity given in \cite{2001ApJ...557..736G} to our desired temperature-dependent mean opacity, we assume that for dust with temperature $T_{\rm gr}$, the dominant contribution to the mean opacity comes from frequencies close to a frequency $\bar{\nu}$ that is given by $h \bar \nu = \alpha k T_{\rm gr}$, where $\alpha = 2.70$. At a reference temperature $T_{0} = 6.75$~K, this procedure yields an opacity
  \begin{equation}\label{goldsmith}
  \begin{array}{rl}
    \kappa(T_0)
     =& 3.3 \times 10^{-26} \alpha (\rm n/2\rho_{\rm gas})\\
                              =& 2.664 \times 10^{-2}/(1 + 4 [\rm He])
   \end{array}
\end{equation} 
where [He] is the helium abundance, and $n$ is the number density of hydrogen nuclei. At other temperatures, $\kappa \propto T_{\rm gr}^{2}$, so long as $T_{\rm gr} < 200$~K. For grain temperatures larger than 200~K, it is necessary to account for the effects of ice-mantle evaporation, while at much higher grain temperatures, the opacity falls off extremely rapidly due to the melting of the grains. We account for these effects \citep[see][]{2003A&A...410..611S} and so our opacity varies with dust temperature following the relationship
\begin{equation}\label{kappa}
\kappa = \kappa(T_0) \times \left\{ 
      \begin{array}{ll}
        T^2 &\hspace{.5in} \mbox{T $<$ 200K} \\
        T^0 &\hspace{.5in} \mbox{200K $<$ T $<$ 1500K}\\
        T^{-12} &\hspace{.5in} \mbox{T $>$ 1500K}
       \end{array} \right.
\end{equation}

\subsection{Setup and Initial conditions}\label{setup}
\begin{center}
\begin{table}[ht]
{\small
\hfill{}
\begin{tabular}{c|rrcc}
\noalign{\smallskip}
\hline \hline
Resolution & Number of & Particle & Turbulence & Angular\\
Level           &  Particles   & Mass     &                       & Momentum\\
\cline{3-5}
\hline
                     &                  & ($10^{-5} \rm M_{\odot}$) & ($E_{\rm turb}/|E_{\rm grav}|$) & ($E_{\rm rot}/|E_{\rm grav}|$)\\
\hline
High & $40 \times 10^6$ &   $2.5$ & 0.1 & 0.02\\
Low  & $4 \times 10^6  $ & $25.0$ & 0.1 & 0.02\\
           &                               &               & 0.0 & 0.00\\
\hline
\end{tabular}}
\hfill{}
\caption{Simulation properties.
}
\label{tsim}
\end{table}
\end{center}
We performed three sets of simulations, two at low resolution and one at high resolution. The details are shown in Table~\ref{tsim}. Our low resolution simulations were performed to explore the thermal evolution of the gas during the collapse, and had 4 million SPH particles which was insufficient to fully resolve fragmentation. We used these simulations to model the collapse of an initially uniform gas cloud with an initial number density of $10^{5} \: {\rm cm^{-3}}$ and an initial temperature of $300 \: {\rm K}$.  We modelled two different metallicities (10$^{-4} {\rm Z}_{\odot}$ and 10$^{-5} {\rm Z}_{\odot}$). The initial cloud mass was $1000 \: {\rm M_{\odot}}$, and the mass resolution was $25 \times 10^{-3} \: {\rm M_{\odot}}$. In one set of low-resolution simulations the gas was initially at rest, while in the other, we included small amounts of turbulent and rotational energy, with $E_{\rm turb}/|E_{\rm grav}|  = 0.1$ and $\beta = E_{\rm rot}/|E_{\rm grav}| = 0.02$, where $E_{\rm grav}$ is the gravitational potential energy, $E_{\rm turb}$ is the turbulent kinetic energy and $E_{\rm rot}$ is the rotational energy. For our high resolution simulations, which were designed to investigate whether the gas would fragment, we employed 40 million SPH particles. We adopted initial conditions similar to those in the low-resolution run with turbulence and rotation. 
Again, we simulated two metallicities, 10$^{-4} {\rm Z}_{\odot}$ and 10$^{-5} {\rm Z}_{\odot}$. The mass resolution (taken to be 100 times the SPH particle mass) was $2.5 \times 10^{-3} {\rm M}_{\odot}$. 

\section{Analysis}\label{anal}

\subsection{Thermodynamical evolution of gas and dust}
\begin{figure}
  \centering
    \includegraphics[width=0.5\textwidth]{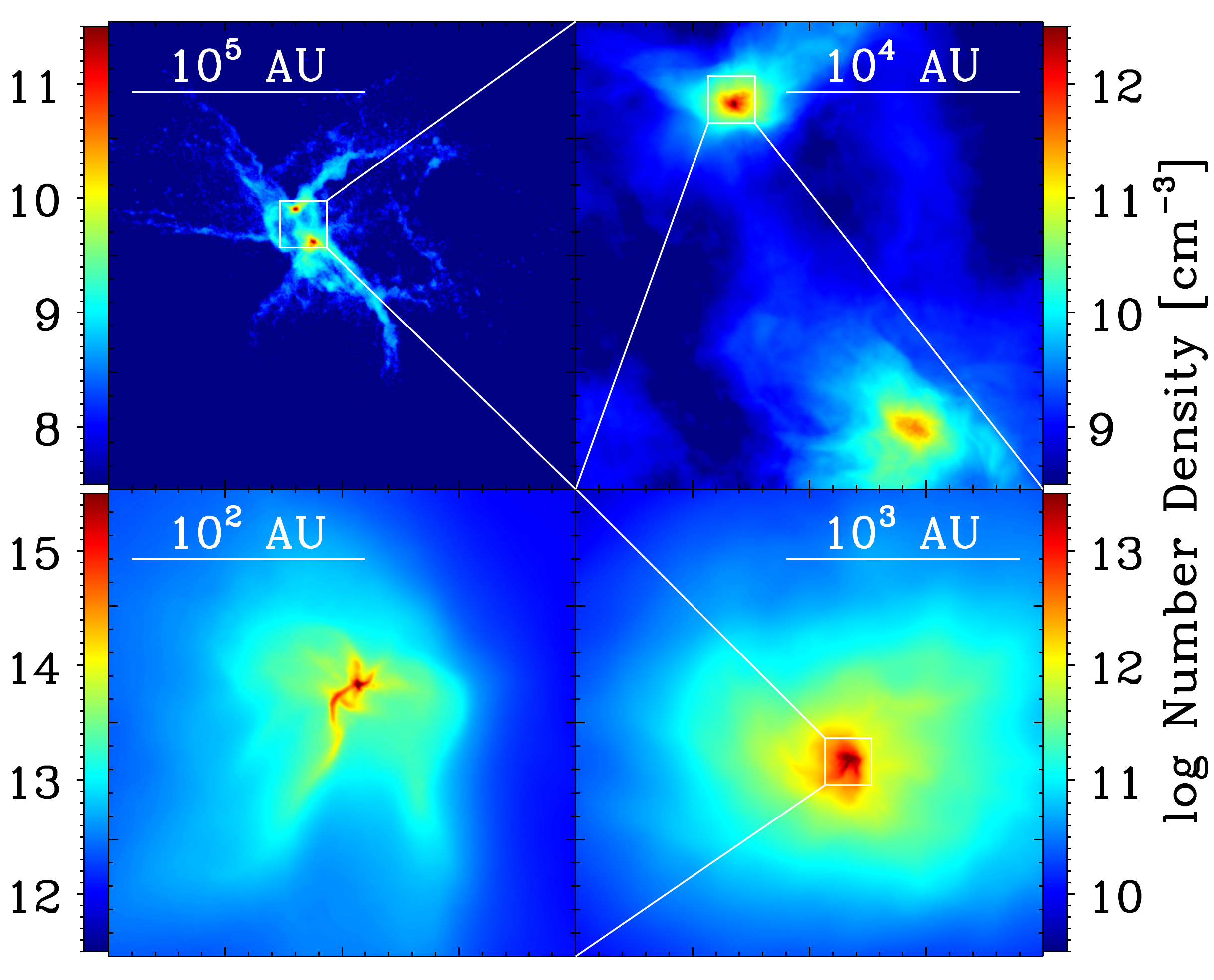}
    \caption{Number density maps for a slice through the high density region. The image shows a sequence of zooms in the density structure in the gas immediately before the formation of the first protostar. \label{nall}}
\end{figure}
In Figure~\ref{nt}, we compare the evolution of the dust and gas temperatures in the low-resolution simulations. The dust temperature, shown in the lower part of the panels, varies from the CMB temperature in the low density region to the gas temperature at much higher densities. 
At densities higher than $10^{11}$--$10^{12} \: \rm cm^{-3}$, dust cooling starts to be effective and begins to cool the gas. The gas temperature decreases to roughly 600~K in the $10^{-5} \: \rm Z_{\odot}$ simulations, and 300~K in the Z $= 10^{-4} \rm Z_{\odot}$ case. This temperature decrease significantly increases the number of Jeans masses present in the collapsing region, making the gas unstable to fragmentation. The dust and the gas temperatures couple for densities higher then $10^{13} \rm cm^{-3}$, when the compressional heating starts to dominate again over the dust cooling. 
The subsequent evolution of the gas is close to adiabatic. If we compare the results of the runs with and without rotation and turbulence, then the most obvious difference is the much greater scatter in the $n-T$ diagram in the former case. Variations in the infall velocity lead to different fluid elements undergoing different amounts of compressional heating. The overall effect is to reduce both the infall velocity and the average compressional heating rate. This allows dust cooling to dominate at a density that is up to five times smaller than in the case without rotation or turbulence. The gas also reaches a lower temperature, cooling down to $\approx$ 200K (instead of 300K) for the Z $= 10^{-4} \rm Z_{\odot}$ case, and to $\approx$ 400K (instead of 600K) for the Z $= 10^{-5} \rm Z_{\odot}$ case. This behavior shows that it is essential to use 3D simulations to follow the evolution of the collapsing gas. A similar effect can be seen in \cite{2011ApJ...727..110C}. 

\subsection{Fragmentation}
\begin{figure}  
  \centering
    \includegraphics[width=0.5\textwidth]{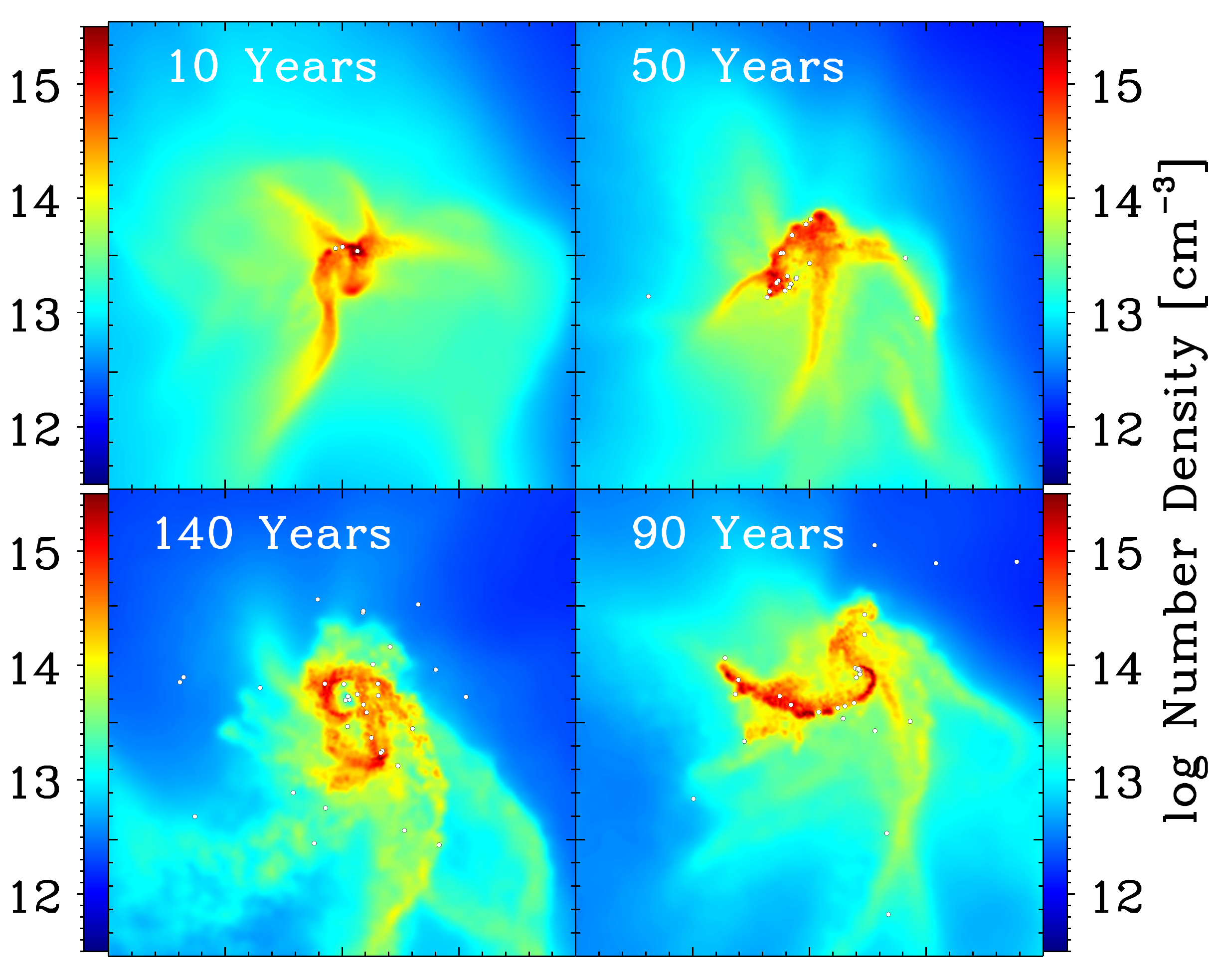}
    \caption{Number density map showing a slice in the densest clump, and the sink formation time evolution, for the 40 million particles simulation, and Z = 10$^{-4}$Z$_{\odot}$. The box is 100AU x 100AU and the time is measured from the formation of the first sink particle.\label{clumpall}}
\end{figure}

We follow the thermodynamical evolution of the gas up to very high densities of order $10^{17} \rm cm^{-3} $, where the Jeans mass is $\approx 10^{-2} \rm M_{\odot}$, and so we need a high resolution simulation to study the fragmentation behaviour.
The transport of angular momentum to smaller scales during the collapse leads to the formation of a dense disk-like structure, supported by rotation which then fragments into several objects.
Figure~\ref{nall} shows the density structure in the gas immediately before the formation of the first protostar. The top-left panel shows a density slice on a scale comparable to the size of the initial gas distribution. The structure is very filamentary and there are two main overdense clumps in the center. If we zoom in on one of the clumps, we see that its internal structure is also filamentary. We can follow the collapse down to scales of the order of an AU, but at this point we reach the limit of our computational approach: as the gas collapses further, the Courant timestep becomes very small, making it difficult to follow the further evolution of the cloud. In order to avoid this difficulty, we replace very dense, gravitationally bound, and collapsing regions by sink particles. 
Once the conditions for sink particle creation are met, they start to form in the highest density regions (Figure \ref{clumpall}). Due to interactions with other sink particles that result in an increase in velocity, some sink particles can be ejected from the high-density region, but most of the particles still remain within the dense gas.  Within 137 years of the formation of the first sink particle, 45 sink particles have formed. At this time, approximately $4.6 \rm M_{\odot}$ of gas has been accreted by the sink particles.

\subsection{Properties of the fragments}
\begin{figure}[H]  
  \centering
    \includegraphics[width=0.5\textwidth]{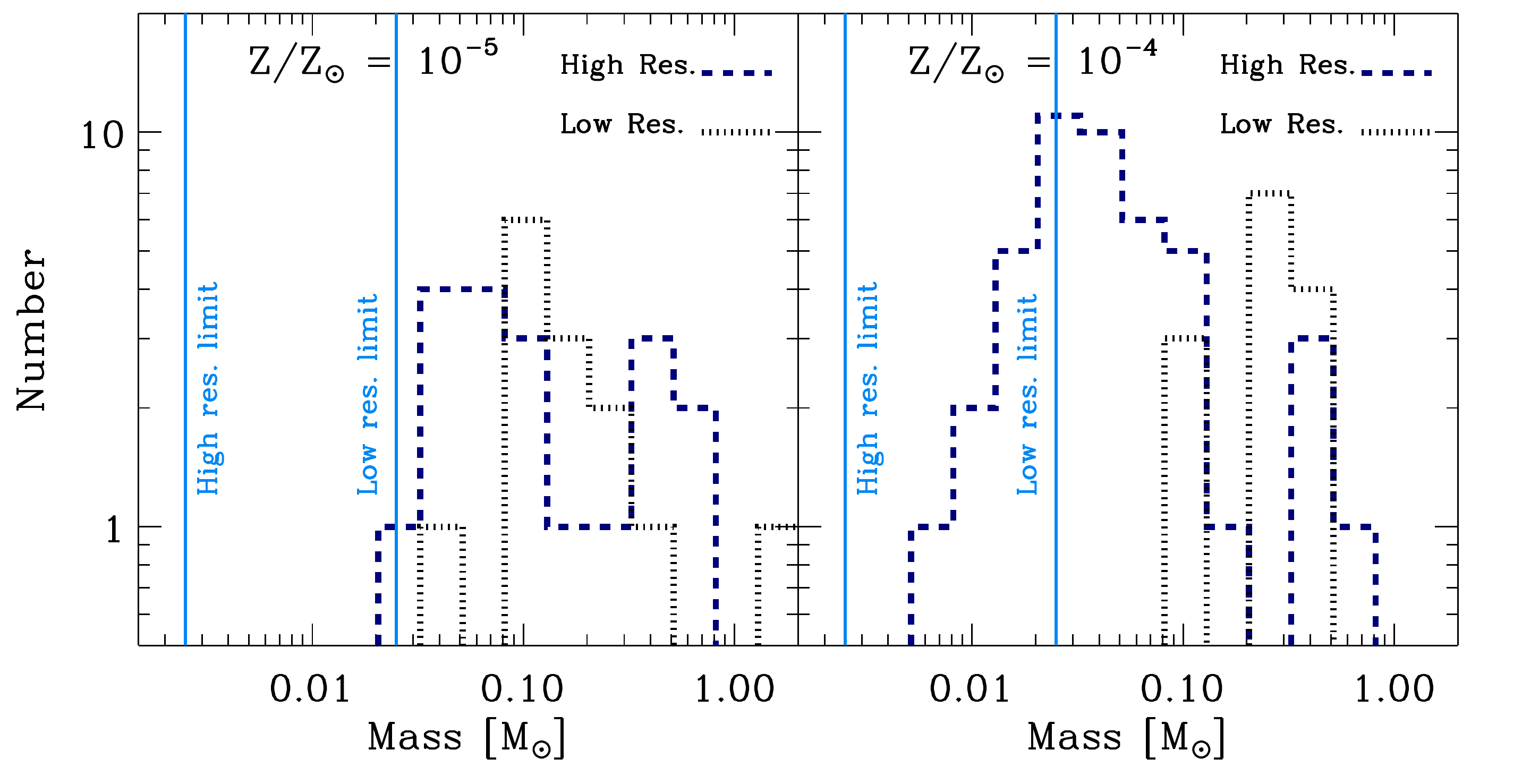}
    \caption{Sink particle mass function at the end of the simulations. High and low resolution results and corresponding resolution limits are shown. To resolve the fragmentation, the mass resolution should be smaller than the Jeans mass at the point in the temperature-density diagram where dust and gas couple and the compressional heating starts to dominate over the dust cooling. At the time shown, around 5 M$_{\odot}$ of gas had been accreted by the sink particles in each simulation.\label{sinkmf}}
\end{figure}

Figure \ref{sinkmf} shows the mass distribution of sink particles when we stop the calculation. We typically find masses below $1 \rm M_{\odot}$, with somewhat smaller values in the $10^{-4} \rm Z _{\odot}$ case compared to the $10^{-5} \rm Z _{\odot}$ case. Both histograms have the lowest sink particle mass well above the resolution limit of $0.0025M_{\odot}$. Note that in both cases, we are still looking at the very early stages of star cluster evolution. As a consequence, the sink particle masses in Figure \ref{sinkmf} are not the same as the final protostellar masses --  there are many mechanisms that will affect the mass function, such as continuing accretion, mergers between the newly formed protostars, feedback from winds, jets and luminosity accretion, etc. Nevertheless, we can speculate that the typical stellar mass is similar to what is observed for Pop II stars in the Milky Way. This suggests that the transition from high-mass primordial stars to Population II stars with mass function similar to that at the present day occurs early in the metal evolution history of the universe, at metallicities $\rm Z_{crit} < 10^{-5}Z_{\odot}$. The number of protostars formed by the end of the simulation, for both $10^{-4} Z_{\odot}$ (45) and $10^{-5} Z_{\odot}$ (19) cases, is much larger than the initial number of Jeans masses (3) in the cloud.

\section{Conclusions}\label{conc}
In this paper we have addressed the question of whether dust cooling can lead to the fragmentation of low-metallicity star-forming clouds. For this purpose we performed numerical simulations 
to follow the thermodynamical and chemical evolution of collapsing clouds. The chemical model included a primordial chemical network together with a description of dust evolution, where the dust temperature was calculated by solving self-consistently the thermal energy equilibrium equation
.\\
We performed three sets of simulations, two at low resolution and one at high resolution (Table \ref{tsim}). All simulations had an initial cloud mass of 1000 M$_{\odot}$, number density of 10$^5$ cm$^{-3}$, and temperature of 300K. We tested two different metallicities (10$^{-4} {\rm Z}_{\odot}$ and 10$^{-5} {\rm Z}_{\odot}$), and also the inclusion of small amounts of turbulent and rotational energies.\\
We found in all simulations that dust can effectively cool the gas, for number densities higher than $10^{11} \rm cm^{-3}$. An increase in metallicity implies a higher dust-to-gas ratio, and consequently stronger cooling by dust. This is reflected in a lower temperature of the dense gas in the higher metallicity simulation.\\
For the low resolution case, we tested the effect of adding turbulence and rotation. These diminish the infall velocity, leading to different fluid elements undergoing different amounts of compressional heating. This lack of heating allows the gas to reach a lower temperature.\\
We found that the transport of angular momentum to smaller scales lead to the formation of a disk-like structure, which then fragmented into a number of low mass objects.\\
We conclude that the dust is already an efficient coolant even at metallicities as low as 10$^{-5}$ or 10$^{-4}Z_{\odot}$, in agreement with previous works \citep{2008ApJ...672..757C, 2010ApJ...722.1793O, 2002ApJ...571...30S, 2006MNRAS.369.1437S, 2006ApJ...642L..61T, 2008ApJ...676L..45T}. Our results support the idea that dust cooling can play an important role in the fragmentation of molecular clouds and the evolution of the stellar IMF.\\
\acknowledgments
{We thank Tom Abel, Volker Brom, Kazuyuki Omukai, Raffaella Schneider, Rowan Smith, and Naoki Yoshida for useful comments. The present work is supported by the \emph{Landesstiftung Baden W\"urttemberg} via their program International Collaboration II (grant P-LS-SPII/18), the German \emph{Bundesministerium f\"ur Bildung und Forschung} via the ASTRONET project STAR FORMAT (grant 05A09VHA), a Frontier grant of Heidelberg University sponsored by the German Excellence Initiative, the International Max Planck Research School for Astronomy and Cosmic Physics at the University of Heidelberg (IMPRS-HD).
All computations described here were performed at the \emph{Leibniz-Rechenzentrum}, National Supercomputer HLRB-II  (\emph{Bayerische Akademie der Wissenschaften}), and on the HPC-GPU Cluster Kolob (University of Heidelberg)
.\\}

\end{document}